\documentclass[runningheads]{llncs}
\usepackage{latexsym}
\usepackage{amsfonts}
\usepackage{amssymb}
\usepackage{amsmath} 
\usepackage{times}

\usepackage[pdftex]{graphicx}

\newcommand{\U}{{\mathcal U}}
\newcommand{\R}{{\mathcal R}}
\newcommand{\Z}{{\mathbb Z}}

\newcommand{\Alw}{{\mathcal Alw }}
\newcommand{\SNC}{{\mathcal S}}
\newcommand{\TRG}{{\mathcal T}}
\newcommand{\Som}{{\mathcal Som }}

\newcommand{\MF}[2]{\langle\langle\text{MF}(#1,#2)\rangle\rangle} 
\newcommand{\MP}[2]{\langle\langle\text{MP}(#1,#2)\rangle\rangle}

\begin{document}

\title{A Metric Encoding for Bounded Model Checking\\(extended version)\thanks{
Work partially supported by the European Commission, Programme
IDEAS-ERC, Project 227977-SMSCom.}}

\author{Matteo Pradella\inst{1} \and Angelo Morzenti\inst{2} \and Pierluigi
San Pietro\inst{2}}

\institute{CNR IEIIT-MI, Milano, Italy %
  \and Dipartimento di Elettronica e Informazione, Politecnico di Milano, Italy \\
  \email{\{pradella, morzenti, sanpietr\}@elet.polimi.it}}

\maketitle

\begin{abstract}
In Bounded Model Checking both the system model and the checked property are translated into a Boolean formula to be analyzed by a SAT-solver. 
We introduce a new encoding technique which is particularly optimized for managing quantitative future and past metric temporal operators, typically found in properties of hard real time systems. The encoding is simple and intuitive in principle, but it is made more complex by the presence, typical of the Bounded Model Checking technique, of backward and forward loops used to represent an ultimately periodic infinite domain by a finite structure. 
We report and comment on the new encoding technique and on an extensive set of experiments carried out to assess its feasibility and effectiveness. 
\end{abstract}

\section{Introduction}\label{introduction}
In Bounded Model Checking \cite{BCCZ99} a system under analysis is
modeled as a finite-state transition system and a property to be
checked is expressed as a formula in temporal logic. The model and the property are
both suitably translated into boolean logic formulae, so that the
model checking problem is expressed as an instance of a SAT problem,
that can be solved efficiently thanks to the significant improvements
that occurred in recent years in the technology of the SAT-solver
tools \cite{Chaff01,ES03}. Infinite,
ultimately periodic temporal structures that assign a value to every
element of the model alphabet are encoded through a finite set of
boolean variables, and the cyclic structure of the time domain is
encoded into a set of {\em loop selector variables} that mark the
start and end points of the period.  As it usually occurs in a model checking
framework, a (bounded) model-checker tool can either prove a property
or disprove it by exhibiting a counter example, thus providing means
to support simulation, test case generation, etc.

In previous work \cite{PMS07}, we introduced techniques for managing
bi-infinite time in bounded model checking, thus allowing for a more
simple and systematic use of past operators in Linear Temporal Logic. 
In 
\cite{ictac08,ase08}, we took advantage of the fact that, in bounded
model-checking, both the model and the formula to be checked are ultimately
translated into boolean logic.
This permits to provide the model not only as a state-transition system, but, alternatively, as a set of temporal logic formulae. We call this a \textit{descriptive} model, as opposed to the term \textit{operational} model used in case it consists of a state-transition system.
The descriptive model is much more readable and concise if the adopted logic includes past and metric temporal operators, allowing for a
great flexibility in the degree of detail and abstraction that the designer can adopt in providing the system model. 
The model-checking problem is reduced to the problem of satisfiability for a boolean formula that encodes both the modeled system and its conjectured property to be verified, hence the name Bounded \emph{Satisfiability} Checking that we adopted for this approach.

In this paper we take a further step to support efficient Bounded Satisfiability- and Bounded Model-checking by introducing a new encoding technique that is particularly efficient in case of temporal logic formulae that contain time constants having a high numerical value. 

In previous approaches \cite{BH+06,PMS07,ictac08,ase08} the operators of temporal logic
that express in a precise and quantitative way some timing constraints were
encoded by (rather inefficiently) translating them into
combinations of non-metric Linear Temporal Logic operators. For instance,
the metric temporal logic formula $\lozenge_{=d} P$, which asserts that
property P holds at $d$ time units in the future (w.r.t the implicit
present time at which the formula is asserted) would be translated into $d$
nested applications of the LTL next-time operator, $\circ^d P$, and then encoded as a series of operator applications, with obvious overhead. 

The new encoding for the metric operators translates the time constants in a way that makes the resulting boolean formula much more compact, and the verification carried out by the SAT solver-based tools significantly faster. 

Thus our technique can be usefully applied to all cases where temporal logic formulae that embed important time constants are used. This is both the case of Bounded Satisfiability Checking, where the system model is expressed as a (typically quite large) set of  metric temporal logic formulae, and also of more traditional Bounded Model Checking, when the model of the system under analysis is provided by means of a state transition system but one intends to check a hard real-time property with explicit, quantitatively stated timing constraints. 

The paper is structured as follows. In Section 2 we provide background and
motivations for our work. Section 3 introduces the new metric encoding and
analyzes its main features and properties. Section 4 provides an assessment
of the new encoding by reporting the experimental results obtained on a set
of significant benchmark case studies. Finally, in Section 5 we draw
conclusions.

\section{Preliminaries}\label{preliminaries}
In this section, to make the paper more readable and self-contained, we provide background material on Metric Temporal Logic and bi-infinite time, on Boundel Model- and Satisfiability-Checking, and on the Zot toolkit.

\subsection{A metric temporal logic on bi-infinite time}

We first recall here Linear Temporal Logic with past operators (PLTL),
in the version introduced by Kamp \cite{K68}, and next extend it with metric temporal operators.

{\bf Syntax of PLTL} The alphabet of PLTL includes: a finite set $Ap$
of propositional letters; two propositional connectives $\neg, \vee$
(from which other traditional connectives such as $\top, \bot, \neg,
\vee, \wedge, \rightarrow, \dots$ may be defined); four temporal
operators (from which other temporal operators can be derived): 
``until''  $\U$,  ``next-time'' $\circ$, 
``since'' $\SNC$ and  ``past-time'' (or Yesterday)
, $\bullet$. Formulae are defined in the usual inductive way:
a propositional letter $p\in Ap$ is a formula; $\neg\phi, \phi \vee
\psi, \phi \U \psi, \circ\phi, \phi \SNC \psi, \bullet\phi$, where
$\phi, \psi$ are formulae; nothing else is a formula.

The traditional ``eventually'' and ``globally'' operators may be defined as: $
\lozenge \phi$ is $\top \U\phi$, $\square \phi$ is $\neg\lozenge\neg
\phi$. Their past counterparts are: $ \blacklozenge \phi$ is $\top
\SNC\phi$, $\blacksquare \phi$ is $\neg\blacklozenge\neg
\phi$. 
Another useful operator for PLTL is  ``Always'' $\Alw$,
defined as $\Alw \; \phi := \Box \phi \land \blacksquare
\phi$. The intended meaning of $\Alw \; \phi$ is that $\phi$ must hold
in every instant in the future and in the past. Its dual is 
``Sometimes'' $\Som \; \phi$ defined as $\neg \Alw \neg \phi$.

The dual operators of Until and Since, i.e.,  ``Release'' 
$\R$: $\phi \R \psi$ is $\neg(\neg \phi \U \neg \psi)$, and, respectively,
 ``Trigger'' $\TRG$: $\phi \TRG \psi$ is $\neg(\neg
\phi \SNC \neg \psi)$, allow 
the convenient {\em
  positive normal form}: 
Formulae are in positive normal form if their alphabet is
$\{\wedge, \vee, \U, \R, \circ,\SNC,\bullet, \TRG \} \cup Ap \cup
\overline{Ap}$, where $\overline{Ap}$ is the set of formulae of the
form $\neg p$ for $p\in Ap$. This form, where negations may only occur on atoms, 
is very convenient when defining encodings of PLTL into
propositional logic. Every PLTL formula $\phi$ on the
alphabet $\{\neg, \vee, \U, \circ,\SNC, \bullet\} \cup Ap$ may be
transformed into an equivalent formula $\phi'$ in positive normal
form.

For the sake of brevity, we also allow
$n$-ary predicate letters (with $n\ge1$) and the $\forall,
\exists$ quantifiers as long as their domains are finite. Hence,
one can write, e.g., formulae of the form: $ \exists p \
\mathrm{gr}(p) $, with $p$ ranging over $\{1,2,3\}$ as a shorthand
for $\bigvee_{p\in\{1,2,3\}} \mathrm{gr}_p$.

{\bf Semantics of PLTL } 
In our past work \cite{PMS07}, we have introduced a variant of bounded model
checking where the underlying, ultimately periodic timing structure was not
bounded to be infinite only in the future, but may extend indefinitely also
towards the past, thus allowing for a simple and intuitive modeling of
continuously functioning systems like monitoring and control devices. In
\cite{ictac08}, we investigated the performance of verification in many
case studies, showing that tool performance on bi-infinite structures is
comparable to that on mono-infinite ones. Hence adopting a bi-infinite
notion of time does not impose very significant penalties to the efficiency
of bounded model checking and bounded satisfiability checking. Therefore, in what follows, we present only the simpler bi-infinite semantics of PLTL.  Each experiment of Section \ref{casestudy} use either bi-infinite time (when there are past operators) or mono-infinite time (typically, when there are only future operators).

A bi-infinite word $S$ over alphabet $2^{Ap}$ (also called
a $\Z$-word) is a function $S:\Z \longrightarrow 2^{Ap}$. Hence,
each position $j$ of $S$, denoted by $S_j$, is in $2^{Ap}$ for every $j$. Word $S$ is also denoted as $\dots
S_{-1}S_0S_{1}\dots$.  The set of all
bi-infinite words over $2^{Ap}$ is denoted by $(2^{Ap})^\Z$. 


For all PLTL formulae $\phi$, for all $S\in (2^{Ap})^\Z$, for all
integer numbers $i$, the satisfaction relation  $S,i\models \phi$ is
defined as follows.

$\begin{array}{l}
S,i\models p,  \Longleftrightarrow p\in S_i, \mathrm{for } \ p\in Ap  \\
S,i\models \neg\phi \Longleftrightarrow S,i\not\models \phi\\
S,i\models \phi\vee\psi \Longleftrightarrow  S,i\models \phi \ \mathrm{or} \ S,i\models \psi\\
S,i\models \circ\phi \Longleftrightarrow  S,i+1\models \phi\\
S,i\models \phi\U\psi \iff
\exists k\ge 0 \mid  S,i+k \models
\psi, \ \mathrm{and} \  S,i+j \models \phi  \ \forall 0\le j< k\\
S,i\models \bullet\phi \Longleftrightarrow S,i-1\models \phi\\
S,i\models \phi\SNC\psi \iff
\exists k\ge 0 \mid S,i-k \models
\psi, \ \mathrm{and} \ S,i-j \models \phi  \ \forall 0\le j< k\\
  \ \\
\end{array}
$

{\bf Metric temporal operators }
Metric operators are
very convenient for modeling hard real time systems, with quantitative time constraints.
The operators introduced in this section do not actually extend the expressive power of PLTL, but 
may lead to more succinct formulae. Their semantics is defined by a straightforward translation $\tau$ 
into PLTL.

Let $\sim \in\{\le, =, \ge \}$), and $c$ be a natural number. 
We consider here two metric operators, one in the future and one in the past: 
the bounded eventually  $ \lozenge_{\sim c} \phi$, and its past counterpart 
$ \blacklozenge_{\sim c} \phi$.
The semantics of the future operators is the following (the past versions
are analogous): 
$$
\begin{array}{l}
    \tau(\lozenge_{= 0} \phi) := \phi \\

    \tau(\lozenge_{= t} \phi) := \circ \tau(\lozenge_{= t-1} \phi),
    \mathrm{ for}\  t > 0 \\

    \hline

    \tau(\lozenge_{\le 0} \phi) := \phi \\

    \tau(\lozenge_{\le t} \phi) :=
    \phi \lor \circ  \tau(\lozenge_{\le t-1} \phi),
    \mathrm{ for}\  t > 0 \\

    \hline

    \tau(\lozenge_{\ge 0} \phi) := \lozenge\phi \\

    \tau(\lozenge_{\ge t} \phi) :=
    \circ \tau(\lozenge_{\ge t-1} \phi),
    \mathrm{ for }\  t > 0 \\

%
%
%
%

\end{array}
$$

Versions of the bounded operators with $\sim \in\{<,>\}$ may be introduced as a shorthand. For instance,
 $\lozenge_{>0} \phi $ stands for $\circ\lozenge_{\ge 0}\phi$.
Other two dual operators are ``bounded globally'': $\square_{\sim c} \phi$ is  $\neg\lozenge_{\sim c}\neg \phi$, 
and its past counterpart is $\blacksquare_{\sim c} \phi$,  which is defined as
$\neg\blacklozenge_{\sim c}\neg \phi$.

Other metric operators are commonly introduced as primitive, such as
bounded versions of $\U$ and $\SNC$ (see e.g. \cite{PMS07}), and then
the bounded eventually operators are derived from them. In our
experience, however, the four operators above are much more common in
specifications, therefore we chose to implement them as native and
leave the others as derived.

Notice that dual w.r.t. negation of metric past  operators, together
with $\bullet$, must be introduced for mono-infinite temporal
structures, to take into account the possibility of referring to
instants outside the temporal domain. In the rest of the paper we
will assume the temporal domain bi-infinite. The complete
mono-infinite encoding is presented in the appendix.

\subsection{Bounded Model Checking vs. Bounded Satisfiability Checking}\label{DvsO}

The traditional approach to verification of finite state systems is based on building an {\em operational model} of the system to be analyzed, 
i.e., a set
of clauses that constrain the transition of the system from a
state valid in one given instant, the {\em current state}, to the
{\em next state}, reached by the modeled system in the successive time
instants.
The property to be
checked, however, is expressed with a different formalism, namely as a formula in temporal logic.
Model checking tools, such as bounded model checkers like SMV, take these two descriptions as input and 
check whether the property is verified on the system, or compute a counterexample.

However, often systems may be described using a complementary style of modeling,  called the {\em descriptive approach}
This is based on the
idea of characterizing the modeled system through its {\em fundamental
properties}, described by means of  temporal logic formulae on an alphabet of
items that correspond to the interface of the system with the
external world, without considering any possible further internal
components that might be necessary for its functioning. Such 
formulae are {\em not} constrained in any way in their form: they may
refer to any time instant, possibly relating actions
and events occurring at any arbitrary distance in time, or they
may constrain values and behaviors for arbitrarily long time
intervals.

Hence, in the descriptive approach both the system under analysis and the property to be checked are expressed in a single uniform notation as formulae of temporal
logic. In this setting, which we called bounded {\em
 satisfiability} checking (BSC \cite{PMS07}), the system under analysis is characterized by a formula $\phi$
(that in all non-trivial cases would be of significant size) and the
additional property to be checked (e.g. a further desired requirement)
is expressed as another (usually much smaller) formula $\psi$. A
bounded model checker in this case is used to prove that any
implementation of the system under analysis possessing the assumed
fundamental properties $\phi$ would also ensure the additional
property $\psi$; in other terms, the model checker would prove that
the formula $\phi \to \psi$ is valid, or equivalently that its
negation is not satisfiable (hence the term \emph{satisfiability checking}).

Satisfiability verification is very useful, in its simplest form, as a
means for performing a sort of testing \cite{FM94} or {\em sanity check} of the specification \cite{MPSS03,RV07}, or to prove properties of correct implementation \cite{ase08} or, more generally, it allows the designer to perform System Requirement Analysis \cite{GM01}. The adoption of a descriptive style in modeling a system under analysis is made possible by the use of Metric Temporal Logic (because formulae might refer to arbitrarily far-away time instant or to arbitrarily long time intervals) and require the adoption of verification methods and tools, like the ones introduced in the present work, that deal efficiently with the important time constants that are typically present in the specification formulae.

\textbf{Example of descriptive vs. operational models: a timed lamp}
As a simplest example of the above introduced concepts we consider the
so-called {\em timer-reset-lamp} (TRL). The lamp has two buttons,
\textit{ON} and \textit{OFF}: when the \textit{ON} button is pressed the
lamp is lighted and it may remain so, if no other event occurs, for
$\Delta$ time units (t.u.), after which it goes off spontaneously. The lighting of the lamp can be terminated by a push of the \emph{OFF} button, or it can be extended by further $\Delta$ t.u. by a new pressure of the \textit{ON} button. To ensure that the pressure of a
button is always meaningful, it is assumed that 
\textit{ON} and \textit{OFF} cannot be pressed simultaneously.


A descriptive model of TRL is based on the following three
propositional letters: \textit{L} (the light is on), \textit{ON} (the button to turn it on is pressed), and \textit{OFF} (the button to turn it off is pressed).
The descriptive model consists of the following axiom:
\[
   \Alw\left(\neg \left(\textit{ON}  \land \textit{OFF}\right) \ \land \
 \left(L \leftrightarrow \exists x
  \left( 0< x \le \Delta \land \blacklozenge_{=x} \textit{ON}\land \neg \blacklozenge_{<x} \textit{OFF} \right)\right)\right)
\]
which expresses the mutual exclusion between the pressing of the
\textit{ON} and \textit{OFF} buttons, and states that the lamp is on (at the current time) if and only
if the \textit{ON} button was pressed not more than $\Delta$ time units ago
and since then the \textit{OFF} button was never pressed. Since the axiom is enclosed in a universal
temporal quantification (an $\Alw$ operator), it must 
hold for all instants of the temporal domain.
This descriptive model, despite its simplicity and succinctness,
characterizes completely the TRL system: starting from it, one can generate 
valid histories for the system, or one can (dis)prove (conjectured) properties.



We now show how an {\em operational model} for the TRL system can
be provided. As mentioned above, the idea is to define, for each
instant, the next system state based on the current state and,
possibly, of the stimuli coming, still at the current time, from
the environment. A brief reflection shows however that the current
state of the TRL system is {\em not} completely characterized by
the value of predicate letter $L$; e.g., if at a given time the lamp is on and no button is pressed, this does \emph{not}
imply that the lamp will still be on at the next
time instant, since this obviously depends on the time that has
elapsed from the last press action on the \textit{ON} button. To
model explicitly this component of the state it is therefore
necessary to introduce a further element in the alphabet of the
model: a counter variable ranging in the interval $[0\dots\Delta]$
to store exactly this information. 

With this addition the definition of the operational model, using any of
the notations adopted in traditional model checkers, like NuSMV or Spin,
becomes an easy exercise, which is not reported here for the sake of
brevity. 

Clearly, an operational model provides a
complete and unambiguous characterization of the TRL system, as
well as the descriptive model.

\subsection{The Zot toolkit}\label{tool}

Zot is an agile and easily extendible bounded model checker, which can
be downloaded at http://home.\-dei.\-polimi.\-it/\-pradella/, together with
the case studies and results described in Section \ref{casestudy}. Zot 
provides a simple language to describe both descriptive and
operational models, and to mix them freely. This is possible since both
models are finally to be translated into boolean logic, to be fed to a SAT
solver (Zot supports various SAT solvers, like  MiniSat \cite{ES03}, and
MiraXT \cite{miraxt}).  The tool supports different logic languages through
a multi-layered approach: its core uses PLTL, and on top of it a decidable
predicative fragment of TRIO \cite{GMM90} is defined (essentially,
equivalent to Metric PLTL).  An interesting feature of Zot is its ability
to support different encodings of temporal logic as SAT problems by means
of plugins. This approach encourages experimentation, as plugins are
expected to be quite simple, compact (usually around 500 lines of code),
easily modifiable, and extendible. 

Zot offers two basic usage modalities:
\begin{enumerate}
\item {\em Bounded satisfiability checking (BSC)}: given as input a
  specification formula, the tool returns a (possibly empty) history
  (i.e., an execution trace of the specified system) which satisfies
  the specification. An empty history means that it is impossible to
  satisfy the specification.

\item {\em Bounded model checking (BMC)}: given as input an
  operational model of the system and a property, the tool returns a (possibly empty)
  history (i.e., an execution trace of the specified system) which
  satisfies it.
\end{enumerate}

The provided output histories have temporal length $\le k$,
the bound \emph{k} being chosen by the user, but may represent infinite behaviors
thanks to the encoding techniques illustrated in Section \ref{encoding}. The BSC/BMC modalities can be used to
check if a property $prop$ of the given specification $spec$ holds over every periodic behavior with period $\le k$. In
this case, the input file contains $spec \land \neg prop$, and, if $prop$ indeed holds, then the
output history is empty. If this is not the case, the output
history is a counterexample, explaining why
$prop$ does not hold.



\section{Encoding of metric temporal logic}\label{encoding}

We  describe next the encoding of PLTL formulae into boolean logic, whose result
includes additional information on the finite structure over which
a formula is interpreted, so that the resulting boolean
formula is satisfied in the finite structure if and only if the original
PLTL formula is satisfied in a (finite or possibly) infinite structure.
For simplicity, we present a variant of the bi-infinite encoding originally
published in \cite{PMS07}, and then introduce metric operators on it.
Indeed, when past operators are introduced over a mono-infinite structure
(e.g., \cite{BH+06}), however, the encoding can be tricky to define,
because of the asymmetric role of future and past: future operators do
extend to infinity, while past operators only deal with a finite prefix.
The reader may refer to \cite{PMS07}, and \cite{ictac08} for a more
thorough comparison between mono- and bi-infinite approaches to bounded
model checking. 


\begin{figure*}

\begin{center}
\includegraphics[width=4.5in]{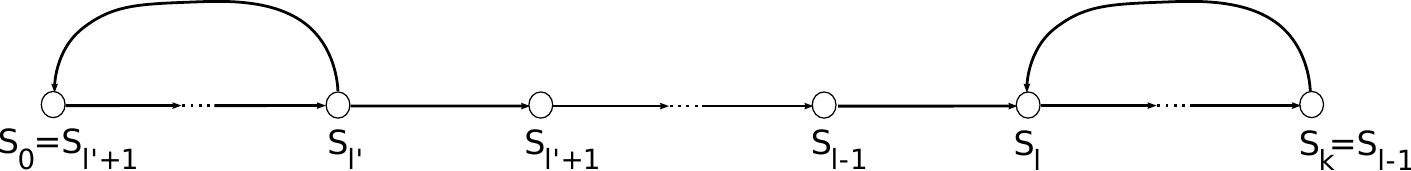}
\caption{A bi-infinite bounded path.}\label{loop-pict}
\end{center}

\end{figure*}
\vspace{-0.5cm}
For brevity in the following we call state $S_i$ the set of
assignments of truth values to propositional variables at time $i$.
The idea on which the encoding is based is graphically depicted in
Figure \ref{loop-pict}. A ultimately periodic bi-infinite
structure has a finite representation that includes a
non periodic portion, and two periodic portions (one towards the future,
and one towards the past). 
 The
interpreter of the formula (in our case, the SAT solver), when it
needs to evaluate a formula at a state beyond the last state
$S_k$, will follow the ``backward link'' and consider the states
$S_l$, $S_{l+1}$, ... as the states following $S_k$.
Analogously,  to evaluate a formula at a state precedent to the first state
$S_0$, it will follow the ``forward link'' and consider the states
$S_l'$, $S_{l'-1}$, ... as the states preceding $S_0$.

The encoding of the model (i.e. the operational description of the system,
if any) is standard - see e.g. \cite{BH+06}. In the following we focus on
the encoding of the logic part $\Phi$ of the system (or its properties).

Let $\Phi$ be a PLTL formula.
Its semantics is given as a set of boolean constraints over the so called
{\em formula variables}, i.e., fresh unconstrained propositional variables.
There is a variable $|[\phi]|_i$ for each subformula $\phi$ of $\Phi$ and
for each instant $0 \le i \le k+1$ (instant $k+1$, which is not explicitly
shown in Figure \ref{loop-pict}, has a particular role in the encoding, as
we will show next).

First, one needs to constrain the propositional operators
in $\Phi$. For instance, if $\phi_1 \land \phi_2$ is a subformula of $\Phi$,
then each variable $|[ \phi_1 \land \phi_2 ]|_i$ must be equivalent to the conjunction of
variables $|[ \phi_1 ]|_i$ and $|[ \phi_2 ]|_i$.

{\em Propositional constraints}, with $p$ denoting a propositional symbol:
\[
\begin{array}{c|c}
    \phi  &  0 \le i \le k \\
    \hline
    p      &
    |[ p ]|_i   \iff  p \in S_i  \\
    \neg p &
    |[ \neg p ]|_i   \iff  p \not\in S_i  \\
    \phi_1 \land \phi_2 &
    |[ \phi_1 \land \phi_2 ]|_i  \iff  |[ \phi_1 ]|_i \land |[ \phi_2 ]|_i \\
    \phi_1 \lor \phi_2 &
    |[ \phi_1 \lor \phi_2 ]|_i   \iff  |[ \phi_1 ]|_i \lor |[ \phi_2 ]|_i \\
\end{array}
\]

The following formulae define the basic temporal behavior of future PLTL
operators, by using their traditional fixpoint characterizations. 


{\em Temporal subformulae constraints}:
\begin{equation}\label{tfc}
\begin{array}{c|c}
    \phi  &  -1 \le i \le k \\
    \hline

    \circ \phi_1 &
    |[ \circ \phi_1 ]|_i \iff |[ \phi_1 ]|_{i+1} \\

    \phi_1 \U \phi_2 &
    |[ \phi_1 \U \phi_2 ]|_i \iff
    |[ \phi_2 ]|_i \lor ( |[ \phi_1 ]|_i \land |[ \phi_1 \U \phi_2
    ]|_{i+1} ) \\

   \phi_1 \R \phi_2 &
   |[ \phi_1 \R \phi_2 ]|_i \iff
   |[ \phi_2 ]|_i \land ( |[ \phi_1 ]|_i \lor |[ \phi_1 \R \phi_2
   ]|_{i+1} ) \\

\end{array}
\end{equation}
\[
\begin{array}{c|c}
    \phi  &  0 \le i \le k+1 \\
    \hline

    \bullet \phi_1 &
    |[ \bullet \phi_1 ]|_i \iff |[ \phi_1 ]|_{i-1} \\

    \phi_1 \SNC \phi_2 &
    |[ \phi_1 \SNC \phi_2 ]|_i \iff
    |[ \phi_2 ]|_i \lor ( |[ \phi_1 ]|_i \land |[ \phi_1 \SNC \phi_2
    ]|_{i-1} ) \\

   \phi_1 \TRG \phi_2 &
   |[ \phi_1 \TRG \phi_2 ]|_i \iff
   |[ \phi_2 ]|_i \land ( |[ \phi_1 ]|_i \lor |[ \phi_1 \TRG \phi_2
   ]|_{i-1} ) \\

\end{array}
\]

Notice that such constraints do not consider the implicit
eventualities that the definitions of $\U$ and $\SNC$ impose (they
treat them as the ``weak'' until and since operators), nor
consider loops in the time structure. 

To deal with eventualities and loops, one has to encode an infinite
structure into a finite one composed of $k+1$ states $S_0,
S_1, \ldots S_k$. The ``future'' loop can be described by means of other 
$k+1$ fresh propositional variables $l_0, l_1, \dots l_k$,
 called {\em loop selector variables}. At most one of these loop selector variables may be
true. If $l_i$ is true then state $S_{i-1}=S_k$, i.e., the bit
vectors representing the state $S_{i-1}$ are identical to those for
state $S_k$. Further propositional variables, $\mathrm{InLoop}_i$ ($0 \le i
\le k $) and $\mathrm{LoopExists}$, respectively  mean that position $i$
is inside a loop and that a loop actually exists in the structure. 
Symmetrically, there are new loop selector variables $l'_i$ to define the
loop which goes towards the past, and the corresponding
propositional letters $\mathrm{InLoop'}_i$, and
$\mathrm{LoopEsists'}$. 

The variables defining the loops are constrained by the following set of formulae.

{\em Loop constraints}:
\[
\begin{array}{c|c}
    \hline
    \mathrm{Base} &
   \begin{array}{c}
    \neg l_0 \land \neg \mathrm{InLoop}_0 \land
    \neg l'_k \land \neg \mathrm{InLoop'}_k
   \end{array}
    \\
    \hline
    1 \le i \le k &
    \begin{array}{c}
        (l_i \Rightarrow S_{i-1} = S_k ) \land
        (\mathrm{InLoop}_i \iff \mathrm{InLoop}_{i-1} \lor l_i) \\
        (\mathrm{InLoop}_{i-1} \Rightarrow \neg l_i) \land
        (\mathrm{LoopExists} \iff \mathrm{InLoop}_k) \\
	\\ 
        (l'_i \Rightarrow S_{i+1} = S_0 ) \land
        (\mathrm{InLoop'}_i \iff \mathrm{InLoop'}_{i+1} \lor l'_i) \\
        (\mathrm{InLoop'}_{i+1} \Rightarrow \neg l'_i) \land
        (\mathrm{LoopExists'} \iff \mathrm{InLoop'}_0)
    \end{array}
\end{array}
\]

The above loop constraints state that the structure may have
at most one loop in the future and at most one loop in the past. 
In the case of a cyclic structure, they allow the
SAT solver to nondeterministically select exactly one of the (possibly) many loops.

To properly define eventualities, we need to
introduce new propositional letters $\langle\langle \diamondsuit
\phi_2 \rangle\rangle_i$, for each $\phi_1 \U \phi_2$ subformula
of $\Phi$, and for every $0 \le i \le k+1$. 
Analogously, we need
to consider subformulae containing the operator $\R$, such as
$\phi_1 \R \phi_2$, by adding the new propositional letters
$\langle\langle \Box \phi_2 \rangle\rangle_i$. This is also symmetrically
applied to $\SNC$ and $\TRG$, using
$\blacklozenge, \blacksquare$. 
Then, constraints
on these eventuality propositions are quite naturally stated as
follows.

{\em Eventuality constraints}:
\[
\begin{array}{c|c}
    \phi & \mathrm{Base} \\
    \hline

    \phi_1 \U \phi_2 &

    \neg \langle\langle \diamondsuit \phi_2 \rangle\rangle_0
    \land
    \left( \mathrm{LoopExists} \Rightarrow \left(
    \begin{array}{c}
        |[ \phi_1 \U \phi_2 ]|_k
        \Rightarrow 
    \langle\langle \diamondsuit \phi_2 \rangle\rangle_k
    \end{array}
    \right) \right) \\

    \phi_1 \R \phi_2 &

   \langle\langle \Box \phi_2 \rangle\rangle_0
   \land
   \left(\mathrm{LoopExists} \Rightarrow \left(
   \begin{array}{c}
   |[ \phi_1 \R \phi_2 ]|_k
       \Leftarrow 
   \langle\langle \Box \phi_2 \rangle\rangle_k
   \end{array}
\right)
\right)\\

   \phi_1 \SNC \phi_2 &

    \neg \langle\langle \blacklozenge \phi_2 \rangle\rangle_k
    \land
    \left(
    \mathrm{LoopExists'} \Rightarrow \left(
    \begin{array}{c}
    |[ \phi_1 \SNC \phi_2 ]|_0 
        \Rightarrow 
    \langle\langle \blacklozenge \phi_2 \rangle\rangle_0
    \end{array}
    \right) \right) \\

   \phi_1 \TRG \phi_2 &

   \langle\langle \blacksquare \phi_2 \rangle\rangle_k
   \land
   \left(\mathrm{LoopExists'} \Rightarrow \left(
       \begin{array}{c}
   |[ \phi_1 \TRG \phi_2 ]|_0 
       \Leftarrow 
   \langle\langle \blacksquare \phi_2 \rangle\rangle_0
   \end{array}
   \right) \right) \\

\end{array}
\]


\[\begin{array}{c|c}
    \phi & 1 \le i \le k  \\
   \hline

    \phi_1 \U \phi_2 &

    \begin{array}{c}
    \langle\langle \diamondsuit \phi_2 \rangle\rangle_i \iff
    \langle\langle \diamondsuit \phi_2 \rangle\rangle_{i-1} \lor
    ( \mathrm{InLoop}_i \land |[ \phi_2 ]|_i )
    \end{array} \\

   \phi_1 \R \phi_2 &

   \begin{array}{c}
   \langle\langle \Box \phi_2 \rangle\rangle_i \iff
   \langle\langle \Box \phi_2 \rangle\rangle_{i-1} \land
   ( \neg \mathrm{InLoop}_i \lor |[ \phi_2 ]|_i ) \\
   \end{array} \\
\end{array}
\]
\[\begin{array}{c|c}
    \phi & 0 \le i \le k-1  \\ 
    \hline 
    \phi_1 \SNC \phi_2  &

    \langle\langle \blacklozenge \phi_2 \rangle\rangle_i \iff
    \langle\langle \blacklozenge \phi_2 \rangle\rangle_{i+1} \lor
    \left(
    \begin{array}{c}
     \mathrm{InLoop'}_i  
     \land  
     |[ \phi_2 ]|_i
    \end{array}
    \right)

    \\

   \phi_1 \TRG \phi_2 &

   \langle\langle \blacksquare \phi_2 \rangle\rangle_i \iff
   \langle\langle \blacksquare \phi_2 \rangle\rangle_{i+1} \land
   \left(
   \begin{array}{c}
    \neg \mathrm{InLoop'}_i 
    \lor  
    |[ \phi_2 ]|_i
   \end{array}
   \right)

\end{array}
\]

The formulae in the following table provide the constraints that must
be included in the encoding, for any subformula $\phi$, to account for
the absence of a forward loop in the structure (the first line of the
table states that if there is no loop nothing is true beyond the
$k$-th state) or its presence (the second line states that if there is
a loop at position $i$ then state $S_{k+1}$ and $S_i$ are equivalent).

{\em Last state constraints}:
\begin{equation}\label{lsc}
\begin{array}{c|c}
    \mathrm{Base} &
    \neg \mathrm{LoopExists} \Rightarrow \neg |[ \phi ]|_{k+1}
    \\
    \hline
    1 \le i \le k &
    l_i \Rightarrow ( |[ \phi ]|_{k+1} \iff |[ \phi ]|_{i})
\end{array}
\end{equation}

Then, symmetrically to the last state, we must define first state
(i.e. 0 time) constraints (notice that in the bi-infinite encoding
instant -1 has a symmetric role of instant $k+1$).

{\em First state constraints}:
\begin{equation}\label{fsc}
\begin{array}{c|c}
    \mathrm{Base} &
    \neg \mathrm{LoopExists'} \Rightarrow \neg |[ \phi ]|_{-1}
    \\
    \hline
    0 \le i \le k-1 &
    l'_i \Rightarrow ( |[ \phi ]|_{-1} \iff |[ \phi ]|_{i})
\end{array}
\end{equation}

The complete encoding of $\Phi$ consists of the logical
conjunction of all above components, together with
$|[\Phi]|_0$ (i.e. $\Phi$ is evaluated only at instant 0).

\subsection{Encoding of the metric operators}\label{metric-stuff}

We present here the additional constraints one has to add to the previous
encoding, to natively support metric operators. 
We actually implemented also a mono-infinite
metric encoding in Zot, but for simplicity we are focusing here only on the
bi-infinite one.

Notice that 
\[
\diamondsuit_{\le t} \phi \iff \neg \Box_{\le t} \neg \phi, 
\qquad 
\diamondsuit_{=t} \phi \iff \Box_{=t} \phi,
\qquad
 \diamondsuit_{\ge t} \phi \iff \diamondsuit_{= t} \diamondsuit \phi
\]
(the past versions are analogous). Hence, in the following we will not
consider the $\Box_{=t}$, 
$\diamondsuit_{\ge t}$, $\Box_{\ge t}$ operators, and their past counterparts.

Ideally, with an {\em unbounded} time structure, the encoding of the
metric operators should be the following one (considering only the future, as
the past is symmetrical):
\[
|[\diamondsuit_{=t}\phi]|_i \iff |[\phi]|_{i+t},
\ \ \ \ \ \ \ \ \ \ \ \   
\ \ \ \ \ \ \ \ \ \ \ \   
|[\Box_{\le t}\phi]|_i \iff \bigwedge_{j=1}^t|[\phi]|_{i+j}
\]

Unfortunately, the presence of a {\em bounded} time structure, in which
bi-infinity is encoded through loops, makes the encoding less
straightforward. 
With simple PLTL one refers at
most to one instant in the future (or in the past) or to an eventuality.
As the reader may notice in the foregoing encoding, 
this is still quite easy, also in the presence of loops.
On the other hand, the presence of metric operators, impacts directly to
the loop-based structure, as logic formulae can now refer to time instants
well beyond a single future (or past) unrolling of the loop.

To represent the values of subformulae inside the future and past loops, we introduce new propositional variables, $\MF{\cdot}{\cdot}$ for the future-tense operators, and $\MP{\cdot}{\cdot}$ for the past ones. For instance, for $\diamondsuit_{=5} \psi$, we introduce $\MF{\psi}{j}$, $0 \le j \le 4$, where the propositions $\MF{\psi}{j}$ are used to represent the value of $\psi$ \emph{j} time units after the starting point of the future loop. This means that, if the future loop selector is at instant 18 (i.e.
$l_{18}$ holds), then $\MF{\psi}{2}$ represents $|[\psi]|_{20}$ (i.e. $\psi$ at instant 18+2).
Analogously and symmetrically, $\MP{\psi}{j}$ are introduced for past
operators with argument $\psi$, and represent the value of $\psi$
\emph{j} time units after the starting point of the past loop. That is, if the past loop
selector is at instant 7 (i.e. $l'_7$), then $\MP{\psi}{2}$ represents
$|[\psi]|_{7-2}$.


The first constraints are introduced for any future or past metric formulae
in $\Phi$.
\begin{equation}\label{mfp}
\begin{array}{c|c}
\phi & 0 \le j \le t-1 \\
	\hline 

    \diamondsuit_{=t} \phi, \  
    \Box_{\le t} \phi,  \ 
    \diamondsuit_{\le t} \phi &
	\MF{\phi}{j} \iff \bigvee_{i=1}^{k} l_i \land
	|[\phi]|_{i+\text{mod}(j,k-i+1)} \\

    \blacklozenge_{=t} \phi,  \ 
    \blacksquare_{\le t} \phi,  \ 
    \blacklozenge_{\le t} \phi  &
	\MP{\phi}{j} \iff \bigvee_{i=0}^{k-1} l'_i \land
	|[\phi]|_{i-\text{mod}(j,i+1)} \\
\end{array}
\end{equation}

We now provide the encoding of every metric operator, composed of two
parts: the first one defines it inside the bounded portion of the
temporal structure (i.e. for instants $i$ in $0 \le i \le k$), and the
other one, based on \emph{MF} and \emph{MP}, for the loop portion.

\begin{equation}\label{enc1}
\begin{array}{c|c}
    \phi  &  -1 \le i \le k \\
    \hline

    \diamondsuit_{=t} \phi &

    |[ \diamondsuit_{=t} \phi ]|_i \iff |[ \phi ]|_{i+t},
    \text{ when } i+t \le k \\
&
    |[ \diamondsuit_{=t} \phi ]|_i \iff 
    \MF{\phi}{t+i-k-1},
    \text{ elsewhere } \\


    \Box_{\le t} \phi &
    \Box_{\le t} \phi \iff
    \bigwedge_{j=1}^{\text{min}(t,k-i)} |[\phi]|_{i+j} 
    \land
    \bigwedge_{j=k+1-i}^{t} \MF{\phi}{i+j-k-1}  \\

    \diamondsuit_{\le t} \phi &
    \diamondsuit_{\le t} \phi \iff
    \bigvee_{j=1}^{\text{min}(t,k-i)} |[\phi]|_{i+j} 
    \lor
    \bigvee_{j=k+1-i}^{t} \MF{\phi}{i+j-k-1}  \\

\end{array}
\end{equation}

\[
\begin{array}{c|c}
 \phi    &  0 \le i \le k+1 \\
    \hline

    \blacklozenge_{=t} \phi &

    |[ \blacklozenge_{=t} \phi ]|_i \iff |[ \phi ]|_{i-t},
    \text{ when } i \ge t \\
&
    |[ \blacklozenge_{=t} \phi ]|_i \iff 
    \MP{\phi}{t-i-1},
    \text{ elsewhere } \\


    \blacksquare_{\le t} \phi &
    \blacksquare_{\le t} \phi \iff
    \bigwedge_{j=1}^{\text{min}(t,i)} |[\phi]|_{i-j} 
    \land
    \bigwedge_{j=i+1}^{t} \MP{\phi}{i+j-1}  \\

    \blacklozenge_{\le t} \phi &
    \blacklozenge_{\le t} \phi \iff
    \bigvee_{j=1}^{\text{min}(t,i)} |[\phi]|_{i-j} 
    \lor
    \bigvee_{j=i+1}^{t} \MP{\phi}{i+j-1}  \\
\end{array}
\]

The most complex part of the metric encoding is the one
considering the behavior on the past loop of future operators, and on the future loop of the past operators.
First, let us consider the behavior of future metric operators on the past loop.

\begin{equation}\label{enc2}
\begin{array}{c|c}
    \phi  &  0 \le i \le k-1 \\
    \hline

    \diamondsuit_{=t} \phi &

    l'_i \Rightarrow
    \left(
    \begin{array}{c}
    \bigwedge_{j=1}^{\text{min}(t,k-i)} 
    (
        |[\phi]|_{i+j} 
	\iff
	|[\phi]|_{\text{mod}(j-1,i+1)}
    ) \land \\

    \bigwedge_{j=k-i+1}^{t}
    \left(
    \begin{array}{c}
    \MF{\phi}{i+j-k-1} 
    \iff \\
	|[\phi]|_{\text{mod}(j-1,i+1)}
    \end{array}
    \right)
  \end{array}
  \right) \\

\\

    \Box_{\le t} \phi &

    \mathrm{InLoop}'_i \Rightarrow
    \left(
    
    |[\Box_{\le t} \phi]|_i \iff
    \left(
    \begin{array}{c}
	    \bigwedge_{j=1}^{\text{min}(k-i,t)}
	    (
	    \neg \mathrm{InLoop}'_{i+j}
\lor 
	    |[\phi]|_{i+j} 
	   ) \land \\
	    \bigwedge_{j=0}^{\text{min}(i,t-1)}
	    (
	    \mathrm{InLoop}'_{\text{min}(k,i+t-j)}
\lor 
	    |[\phi]|_j 
	    ) 
    \end{array}
    \right)
    \right)

\\

    \diamondsuit_{\le t} \phi &

    \mathrm{InLoop}'_i \Rightarrow
    \left(
    
    |[\diamondsuit_{\le t} \phi]|_i \iff
    \left(
    \begin{array}{c}
	    \bigvee_{j=1}^{\text{min}(k-i,t)}
	    (
	    \mathrm{InLoop}'_{i+j}
\land 
	    |[\phi]|_{i+j} 
	   ) \lor \\
	    \bigvee_{j=0}^{\text{min}(i,t-1)}
	    (
	    \neg \mathrm{InLoop}'_{\text{min}(k,i+t-j)}
\land 
	    |[\phi]|_j 
	    ) 
    \end{array}
    \right)
    \right)
\end{array}
\end{equation}

The main aspect to consider is the fact that, if $l'_i$ (i.e. the past loop
selector variable holds at instant $i$), then $i$ has two possible
successors: $i+1$ and 0.
Therefore, if $\diamondsuit_{=4} \phi$ holds at $i$ (which is inside the
past loop), then $\phi$ must hold both at $i+4$, and at $3$.
 This kind of constraint is captured by the upper formula for
 $\diamondsuit_{=t} \phi$, which 
 relates the truth
 values of $\phi$ in instants outside of the past loop (i.e.,
$|[\phi]|_{i+j}$) with the instants inside (i.e.,
 $|[\phi]|_{\text{mod}(j-1,i+1)}$ represents the value of $\phi$ at instants
 going from 0 to $i$, if $l'_i$ holds).  

Another aspect to consider is related to the size of the time constant used (i.e. $t$
in this case). Indeed, if $i+t > k$, then we are considering the behavior
of $\phi$ outside the bound $0..k$. This means that we need to consider the
behavior of $\phi$ also in the future loop, hence we refer to 
    $\MF{\phi}{i+j-k-1}$ (see the lower formula for
$\diamondsuit_{=t} \phi$).

As far as $\Box_{\le t} \phi$  is concerned, 
its behavior inside the past loop is in general expressed by two parts.
The first one considers $\phi$ inside the past loop, starting from
instant $i$ and going forward, towards the right end of the loop (i.e. where
$l'$ holds, say $i'$). This situation is covered by the upper formula for
 $\Box_{\le t} \phi$. If $i+t$ is still inside the past loop (i.e. $i+t \le
i'$), this suffices.
If this is not the case, we must consider the remaining instants, going
from $i'+1$ to $i+t$. Because we are considering the behavior {\em inside} the
past loop, the instant after $i'$ is $0$, so we must translate instants
outside of the loop (i.e. where $InLoop'$ does not hold), to instants going
from 0 to $i+t-i'-1$: in all these instants $\phi$ must hold.
 This constraint is given by the lower formula for 
 $\Box_{\le t} \phi$.

The encoding for the past operators is symmetrical, and is the
following:
\[
\begin{array}{c|c}
     &  1 \le i \le k \\
    \hline

    \blacklozenge_{=t} \phi &

    l_i \Rightarrow
    \left(
    \begin{array}{c}
    \bigwedge_{j=2}^{\text{min}(t,i)} 
    (
        |[\phi]|_{i-j} 
	\iff
	|[\phi]|_{k-\text{mod}(j-1,k-i+1)}
    ) \land \\

    \bigwedge_{j=1+i}^{t}
    \left(
    \begin{array}{c}
    \MP{\phi}{j-i-1} 
    \iff \\
    |[\phi]|_{k-\text{mod}(j-1,k-i+1)}
    \end{array}
    \right)
  \end{array} 
  \right) \\

\\

   \blacksquare_{\le t} \phi &

   \mathrm{InLoop}_i \Rightarrow
    \left(
    
    |[\blacksquare_{\le t} \phi]|_i \iff
\left(
    \begin{array}{c}
	    \bigwedge_{j=1}^{\text{min}(i,t)}
	    (
	    \neg \mathrm{InLoop}_{i-j}
\lor 
	    |[\phi]|_{i-j}
	   ) \land \\
	    \bigwedge_{j=0}^{\text{min}(k-i,t-1)}
	    (
	    \mathrm{InLoop}_{\text{max}(0,i-t+j)}
\lor 
	    |[\phi]|_{k-j} 
	    ) 
    \end{array}

    \right)
    \right)
\\

  \blacklozenge_{\le t} \phi &

   \mathrm{InLoop}_i \Rightarrow
    \left(
    
    |[\blacklozenge_{\le t} \phi]|_i \iff
\left(
    \begin{array}{c}
	    \bigvee_{j=1}^{\text{min}(i,t)}
	    (
	    \mathrm{InLoop}_{i-j}
\land 
	    |[\phi]|_{i-j}
	   ) \lor \\
	    \bigvee_{j=0}^{\text{min}(k-i,t-1)}
	    (
	    \neg  \mathrm{InLoop}_{\text{max}(0,i-t+j)}
\land 
	    |[\phi]|_{k-j} 
	    ) 
    \end{array}

    \right)
    \right)

\end{array}
\]

The actual implementation of the metric encoding contains some
optimizations, not reported here for the sake of brevity, like the re-use,
whenever possible, of the various $\MF{\cdot}{\cdot}$, and
$\MP{\cdot}{\cdot}$ propositional letters.

\medskip

{\bf A first assessment of the encoding} The behavior of the new
encoding has been first experimented on a very simple specification of
a synchronous shift-register, where, at each clock tick, an input bit
is shifted of one position to the right.  A specification of this
system can be described by the following formula:
\[ Alw(in \leftrightarrow \diamondsuit_{=d}out) \]
where $in$ is true when a bit enters the shift register, $out$ is true when a bit ``exits'' the register after a delay $d$ (a constant representing the number of memory bits in the register).
The Zot toolkit has been applied to this simple specification, using the
nonmetric,
PLTL-only encoding (i.e. the one presented in \cite{PMS07}) and the
new metric encoding.

The implemented nonmetric encoding is the one presented in
the current section, without the metric part of Sub-section \ref{metric-stuff}.
In practice, this means that every metric temporal operator is
translated into PLTL before applying the encoding, by means of its
definition of Section \ref{preliminaries}.

\begin{figure}
\begin{center}
\includegraphics[width=14cm]{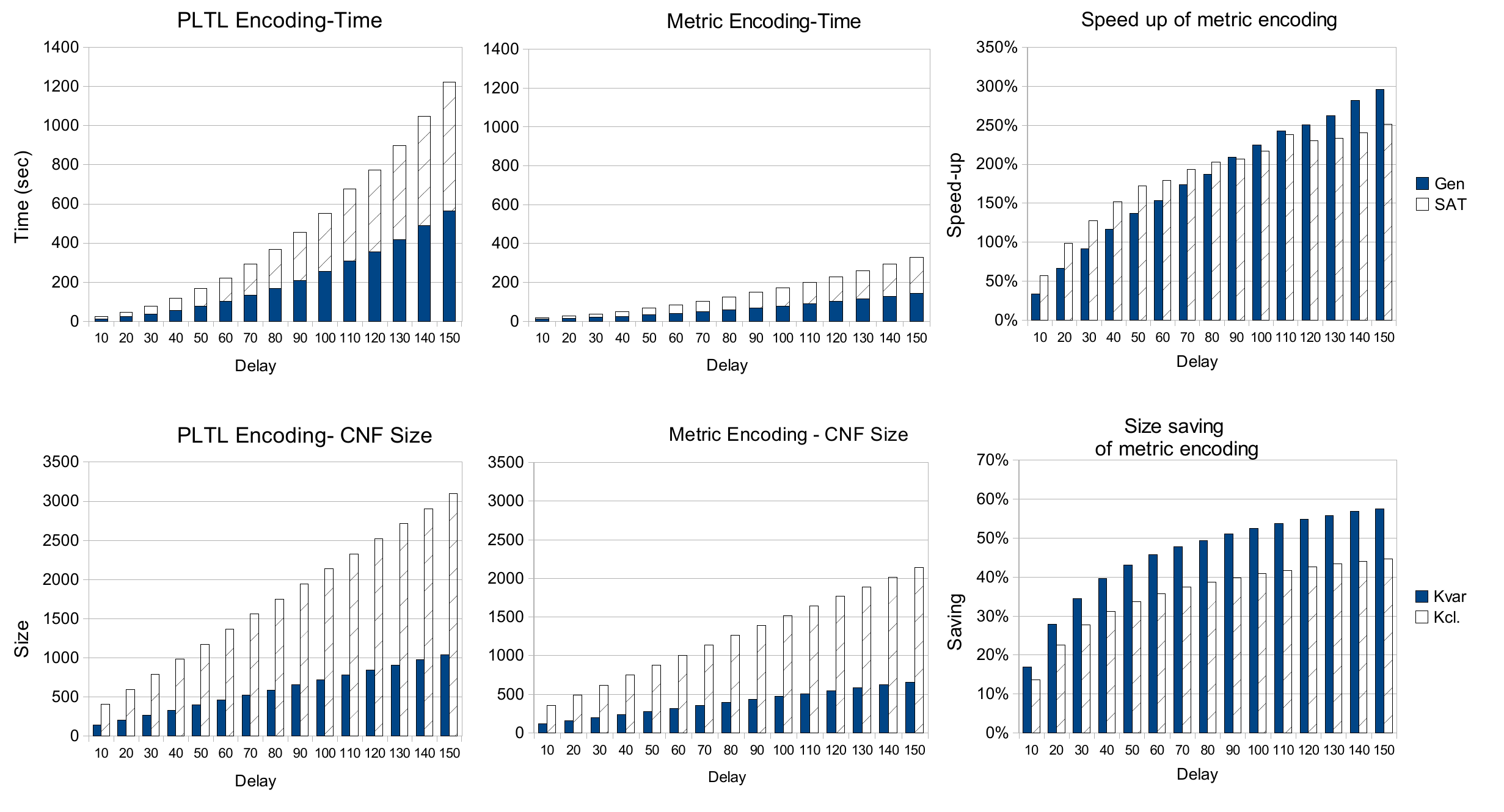}
\caption{Summary of experimental data for the synchronous version of a
Shift Register.}
\label{shiftSync}
\end{center}
\end{figure}

The experimental results (with
the same hardware and software setup described in Section \ref{results}
are graphically shown in Figure~\ref{shiftSync}, where Gen represents the generation phase, i.e., the generation of a boolean formula in conjunctive normal form, starting from the above specification, and SAT represents  the verification phase, performed by a SAT solver, with a bound $k=400$ and various values of delay $d$ (from 10 to 150). 
The first two upper diagrams show the time, in seconds, for Gen and SAT phases, using either a PLTL encoding or the metric encoding, as a function of delay $d$, while the third upper diagram shows the speedup, as a percentage of speed increase over the PLTL encoding, when using the metric encoding, again as a function of delay $d$. 
As one can see, the speed up obtained for both the Gen and SAT phases is proportional to delay $d$, and can be quite substantial (up to 250\% for SAT and 300\% for Gen phases). 
The three lower diagrams report, in a similar way, on the size of the generated boolean formula, in terms of the thousands of variables (Kvar) and clauses (Kcl): the reduction in the size of the generated encoding increases with the value of $d$ and tends to reach a stable value around 60\%.

These results can be explained by comparing the two encodings. In the previous, non-metric encoding the formula $\diamondsuit_{=d}out$ is translated into $d$ nested applications of the next-time operator, $\circ^{d} out$, hence there are $d+1$ subformulae, $\circ^{i} out$ for $0 \leq i \leq d$. For each of these the encoding procedure generates $k+2$ boolean variables, $k$ last state constraints of type (\ref{lsc}), $k$ first state constraints of type (\ref{fsc}), and $k+2$ temporal subformulae constraints of type (\ref{tfc}) for a total of $(d+1) \cdot (k+2)$ variables and $(d+1) \cdot (3 \cdot k + 2)$ constraints. In summary, in the nonmetric encoding we have $O(d \cdot k)$ variables and $O(d \cdot k)$ constraints.  
On the contrary, in the metric encoding of $\diamondsuit_{=d}out$ there are only two subformulae, $\diamondsuit_{=d}out$ itself and $out$. Now the encoding procedure generates $2 \cdot (k+2)$ variables, plus $2 \cdot d$ \textit{MF} variables (see equation \ref{mfp}), for a total of $2 \cdot (d + k + 2)$ variables. It also generates $4 \cdot k$ first and last state constraints of type (\ref{lsc}) and (\ref{fsc}), $k$ constraints of type (\ref{enc1}) plus $d$ constraints of type (\ref{mfp}), each of these having size $O(k)$, and $k$ constraints of type (\ref{enc2}) having size $O(d)$; overall, we have therefore $4 \cdot k + d$ constraints, and their total size is significantly smaller that in the nonmetric encoding, though it is still $O(d \cdot k)$. Thus in the metric encoding we have $O(d + k)$ (less than in the nonmetric case) variables and $O(d \cdot k)$ constraints (same as in the nonmetric case but with a smaller constant factor). The analysis of the other metric temporal operators, $\Box_{\le t} \phi$ and $\diamondsuit_{\le t} \phi$, leads to similar conclusions.


\section{Experimental results}\label{casestudy}

First we briefly describe the five case studies that we adopted for our experiments. 
For all of them we provide both a descriptive and an operational model. 
A complete archive with the files used for the
experiments, and the details of the outcomes, can be found in the Zot web page at
http://home.dei.polimi.it/pradella/.

\textbf{Real-time allocator (RTA)}
This case study, described in \cite{ase08}, consists of a real-time allocator which serves a set of client processes, competing for a shared resource.  
The system numeric parameters are the number of processes $n_{p}$ and the constants $T_{req}$ within which the allocator must respond to the requests, and the maximum time $T_{rel}$ that a process can keep the resource before releasing it. 
In our experiments, both a descriptive and an operational model were considered, using three processes, and with two different system settings for each version: 
a first one with $T_{rel} = T_{req} = 3$, and a second one with $T_{rel} = T_{req} = 10$. We first generated a 
simple run of the system (Property Sat); then we considered four hard real time properties, described in \cite{ase08}, called \emph{Simple Fairness}, \emph{Conditional Fairness}, \emph{Precedence}, and \emph{Suspend Fairness}. It is worth noticing that the formula specifying \emph{Suspend Fairness} includes a relatively high time constant ($T_{rel} \cdot n_p$) and is therefore likely to benefit from the 
metric encoding. 
We adopted the bi-infinite encoding for this case study,  which allowed to consider only regime behaviors, thus abstracting away system initialization.

\textbf{Fischer's protocol (FP)}
FP \cite{Fischer} is a timed mutual exclusion algorithm that allows a
number of timed processes to access a shared resource. 
We considered the system in two variants: one with 3 processes and a delay 5
t.u.; the other one with 4 processes and a delay of 10 t.u.
We used the tool to check the safety property (i.e. it is never possible that two different processes enter their critical sections at the same time instant) 
and to generate a behavior in which there is always at least one alive process.
We adopted  the bi-infinite encoding, for reasons similar to those already explained for RTA case study.

\textbf{Kernel Railway Crossing (KRC)}
This is a standard benchmark in real time
systems verification \cite{HM96}, which we used and described in a previous work \cite{ase08}.
In our example we adopted a descriptive model and studied the KRC problem with two sets of time constants, 
allowing a high degree of nondeterminism on train behavior. In particular,
the first set of constants was: $d_{Max}=9$ and $d_{min}=5$ t.u. for the 
maximum and minimum time for a train to reach the critical region, 
$h_{Max}=6$ and  $h_{min}=3$ for the maximum and minimum time for a train to 
enter the critical region once it is first sensed, and $\gamma=3$ 
for the movement of the bar from up to down and vice versa. The set of time constants for the second experiment was $d_{Max}=19$, $d_{min}=15$, 
$h_{Max}=16$, $h_{min}=13$, and $\gamma=10$. For each of the two settings we proved both satisfiability of the specification (Sat) and the safety property, using a mono-infinite encoding.

\textbf{Timer Reset Lamp (TRL)}
This is the Timer Reset Lamp first presented in \cite{ase08}, with three settings ($\Delta=10$, $\Delta=15$, and $\Delta=20$) and two analyzed
properties (the first one, that the lamp is never lighted for more than
$\Delta$ t.u.: it is false, and the tool generates a counter-example; the second one, namely that the lamp can remain lighted for more than $\Delta$ t.u. 
only if the \emph{ON} button is pushed twice within $\Delta$ t.u., is true). This system was analyzed with a bi-infinite encoding.

\textbf{Asynchronous Shift Register (ASR)}
The simplest case study is an \emph{asynchronous} version of the
Shift Register example discussed in Section \ref{encoding}, where the shift
does not occur at every tick of the clock, but only at a special, completely
asynchronous  \emph{Shift} command. We consider two cases, with the number of bits $n=16$ and 
$n=24$, and we prove satisfiability of the specification and one timed
property (if the \emph{Shift} signal remains true for \emph{n} time units
(t.u.) then the value \emph{In} which was inserted in the Shift register at the beginning of 
the time interval will appear at the opposite side of the register at the end of the time interval). 
This case study was analyzed with reference to a bi-infinite encoding.

\subsection{Results}\label{results}

\begin{figure*}
\begin{center}
\includegraphics[width=14cm]{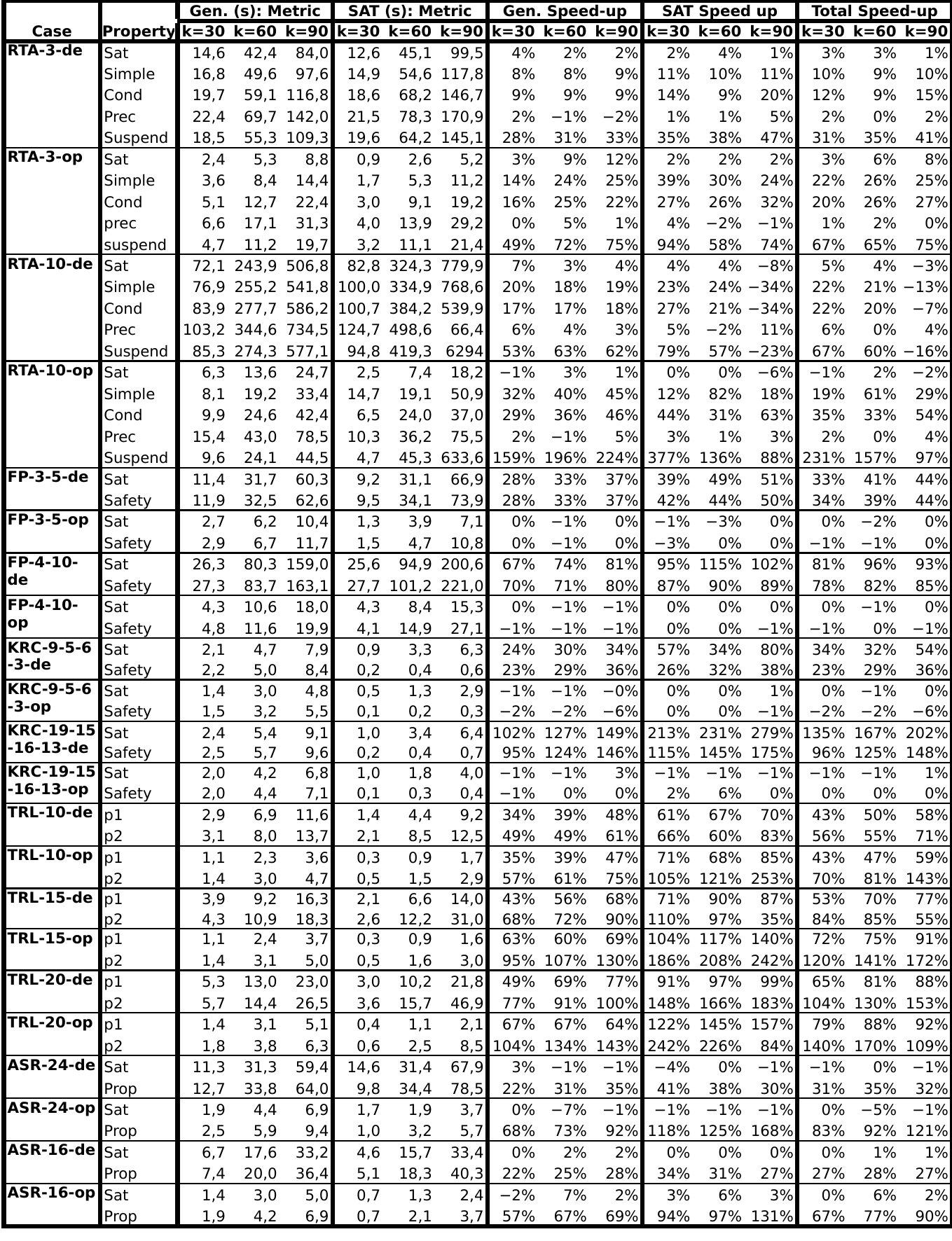}
{\bf Table 1.} Summary of collected experimental data. 
\label{rawData}
\end{center}
\end{figure*}

The experiments were run on a PC equipped with
two XEON 5335 (Quadcore) processors at 2.0 Ghz, with 16 GB RAM, running under Gentoo X86-64 (2008.0). The SAT-solver was MiniSat.
The experimental results are shown in Table 1. The suffix {\em -de} indicates analysis carried out on the descriptive
version of the model, while {\em -op} is used for the operational
version. The table reports, for various values of the bound $k$ (30, 60, and 90), both Generation time, i.e., 
the time in seconds taken for building the encoding and transforming it into 
conjunctive normal form,  and SAT time, i.e., the time in seconds taken by the SAT solver to answer.
Only the timings of the metric version is reported, since the ones of the non-metric version can be obtained by the following speed up measures. 
Performance is gauged by providing three measures of speed up as a percentage of the time taken by the metric version 
(e.g., 0\% means no speed-up, 100\% means double speed, i.e., the encoding is twice as fast, etc.): $\frac{T_\texttt{PLTL} - T_\texttt{metric}}{T_\texttt{metric}}$, 
where $T_\texttt{metric}$ and $T_\texttt{PLTL}$ represent
the time taken by the metric and the PLTL encodings, respectively. 
The first measure shows the speed up in the generation phase, the second in SAT time and the third one in  Total time 
(i.e., in the sum of  Gen and SAT time). 
On average, the speed up is 42,2\% for Gen and 62,2\% for SAT, allowing for a 47,9\% speed up in the total time. 
The best results give speed up of, respectively, 224\%, 377\% and 231\%, while the worst results are -7\%, -34\% and -16\%.

Speed up for SAT time appears to be more variable and less 
predictable than the one for Gen time, although often significantly larger. This is likely caused by the complex and 
involved ways in which the SAT algorithm is influenced by the numerical values of the \emph{k} bound, of the time constants in the 
specification formulae and by their interaction, due to the heuristics that it incorporates. 
For instance, the speed up for Gen increases very regularly with the bound $k$, because of
the smaller size of the formula to be generated, while SAT may vary unpredictably and significantly with
the value of $k$ (e.g., compare property op-P2 for TRL-10, when the speed up increases with $k$, and TRL-20, 
when the speed up actually decreases with $k$).
A thorough discussion of these aspects is out of the scope of the present paper, 
also because they may change from one SAT-solver to another one.


It is easy to realize, as already noticed in Section \ref{encoding} for the example of the synchronous 
shift register, that significant improvements are obtained, with the new
metric encoding, for analysing  Metric 
temporal logic properties with time constants having a fairly high numerical value. The larger the value, 
the larger the speed up.
This is particularly clear for TRL, RTA and FP case studies. 

The fact that the underlying model was descriptive or operational may have a significant impact on verification speed, 
but considering only the speed up the results are much more mixed. For instance, 
the operational versions of FP and KRC, although more efficient, had a worse speed up than their corresponding descriptive cases, 
while the reverse occurred for the operational versions of RTA, ASR and TRL. The only exception is for the Sat case, where no property is checked against the model, and hence
no gain can be obtained for the operational model. 
A decrease in benefit for certain descriptive models may be caused by cases where subformulae 
in metric temporal logic with large time constants are combined with other 
non-metric subformulae.

The measure of the size of the generated formulae is not reported here, but it is worth pointing 
out that, thanks to the new metric encoding, the size is dramatically reduced when there are high 
time constants and/or large $k$ bounds. 
In fact, in the previous, non-metric encoding, size is 
proportional to the \emph{product} of the \emph{k} bound and the numerical value 
of time constants, while in the new, metric encoding size is only proportional to their
\emph{sum}.

\section{Conclusions}
In this paper, a new encoding technique of linear temporal logic into boolean logic is introduced, particularly optimized for managing quantitative future and past metric temporal operators. 
The encoding is simple and intuitive in principle, but it is made more complex by the presence, typical of the technique, of backward and forward loops used to represent an ultimately periodic infinite domain by a finite structure. 

We have shown that, for formulae that include an explicit time constant, like e.g.,  $\lozenge_{= t} \phi$, the new metric encoding permits an improvement, in the size of the generated SAT formula and in the SAT solving time, that is proportional to the numerical value of the time constant. In practical examples, the overall performance improvement is limited by other components of the encoding algorithm that are not related with the value of the time constants (namely, those that encode the structure of the time domain, or the non-metric operators). Therefore, the gain in performance can be reduced in the less favorable cases in which the analyzed formula contains few or no metric temporal operators, or the numerical value of the time constants is quite limited.
 
An extensive set of experiments has been carried out to asses its feasibility and effectiveness for Bounded Model Checking (and Bounded Satisfiability Checking). 
Average speed up in SAT solving time was 62\%.
The experimental results show that the new metric encoding can successfully be 
 applied when the property to analyze includes time constants 
with a fairly high numerical value.

{\bf Acknowledgements:} 
We thank Davide Casiraghi for his valuable
work on Zot's metric plugins.

\bibliographystyle{abbrv}
\bibliography{triobib,automatabib}

\newpage
\section*{Appendix: a Mono-infinite Encoding}\label{mono-enc}

In some sense, the mono-infinite encoding of PLTL is simpler, since
there is only the forward loop to be taken into account. On the other
hand, being the temporal structure mono-infinite, it is possible to
refer to time instants before 0 (e.g. by using $\bullet$ at instant
0).
The typical approach (see e.g. \cite{BH+06}) is to use a default
value for operators referring to instants outside the temporal
domain: in our case, $\bullet \phi$ at 0 is false for any $\phi$.
Because of this, it is necessary to introduce a dual operator for
representing the negation of $\bullet$, which we will denote by $\bullet'$.
Its semantics is given by the following formula:
\[
\bullet \phi \iff \neg \bullet'  \neg \phi.
\]

Next, we present the constraints of Section \ref{encoding}, modified
for a mono-infinite time structure.

\noindent {\em Propositional constraints}, with $p$ denoting a propositional symbol:
\[
\begin{array}{c|c}
    \phi  &  0 \le i \le k \\
    \hline
    p      &
    |[ p ]|_i   \iff  p \in S_i  \\
    \neg p &
    |[ \neg p ]|_i   \iff  p \not\in S_i  \\
    \phi_1 \land \phi_2 &
    |[ \phi_1 \land \phi_2 ]|_i  \iff  |[ \phi_1 ]|_i \land |[ \phi_2 ]|_i \\
    \phi_1 \lor \phi_2 &
    |[ \phi_1 \lor \phi_2 ]|_i   \iff  |[ \phi_1 ]|_i \lor |[ \phi_2 ]|_i \\
\end{array}
\]

\noindent {\em Temporal subformulae constraints}:
\[
\begin{array}{c|c}
    \phi  &  0 \le i \le k \\
    \hline

    \circ \phi_1 &
    |[ \circ \phi_1 ]|_i \iff |[ \phi_1 ]|_{i+1} \\

    \phi_1 \U \phi_2 &
    |[ \phi_1 \U \phi_2 ]|_i \iff
    |[ \phi_2 ]|_i \lor ( |[ \phi_1 ]|_i \land |[ \phi_1 \U \phi_2
    ]|_{i+1} ) \\

   \phi_1 \R \phi_2 &
   |[ \phi_1 \R \phi_2 ]|_i \iff
   |[ \phi_2 ]|_i \land ( |[ \phi_1 ]|_i \lor |[ \phi_1 \R \phi_2
   ]|_{i+1} ) \\

\end{array}
\]
\[
\begin{array}{c|c}
    \phi  &  1 \le i \le k+1 \\
    \hline

    \bullet \phi_1 &
    |[ \bullet \phi_1 ]|_i \iff |[ \phi_1 ]|_{i-1} \\

\bullet' \phi_1 &
    |[ \bullet' \phi_1 ]|_i \iff |[ \phi_1 ]|_{i-1} \\

    \phi_1 \SNC \phi_2 &
    |[ \phi_1 \SNC \phi_2 ]|_i \iff
    |[ \phi_2 ]|_i \lor ( |[ \phi_1 ]|_i \land |[ \phi_1 \SNC \phi_2
    ]|_{i-1} ) \\

   \phi_1 \TRG \phi_2 &
   |[ \phi_1 \TRG \phi_2 ]|_i \iff
   |[ \phi_2 ]|_i \land ( |[ \phi_1 ]|_i \lor |[ \phi_1 \TRG \phi_2
   ]|_{i-1} ) \\

\end{array}
\]

{\em Loop constraints}:
\[
\begin{array}{c|c}
    \hline
    \mathrm{Base} &
   \begin{array}{c}
    \neg l_0 \land \neg \mathrm{InLoop}_0 
   \end{array}
    \\
    \hline
    1 \le i \le k &
    \begin{array}{c}
        (l_i \Rightarrow S_{i-1} = S_k ) \land
        (\mathrm{InLoop}_i \iff \mathrm{InLoop}_{i-1} \lor l_i) \\
        (\mathrm{InLoop}_{i-1} \Rightarrow \neg l_i) \land
        (\mathrm{LoopExists} \iff \mathrm{InLoop}_k) \\
    \end{array}
\end{array}
\]

\noindent {\em Eventuality constraints}:
\[
\begin{array}{c|c}
    \phi & \mathrm{Base} \\
    \hline

    \phi_1 \U \phi_2 &

    \neg \langle\langle \diamondsuit \phi_2 \rangle\rangle_0
    \land
    \left( \mathrm{LoopExists} \Rightarrow \left(
    \begin{array}{c}
        |[ \phi_1 \U \phi_2 ]|_k
        \Rightarrow 
    \langle\langle \diamondsuit \phi_2 \rangle\rangle_k
    \end{array}
    \right) \right) \\

    \phi_1 \R \phi_2 &

   \langle\langle \Box \phi_2 \rangle\rangle_0
   \land
   \left(\mathrm{LoopExists} \Rightarrow \left(
   \begin{array}{c}
   |[ \phi_1 \R \phi_2 ]|_k
       \Leftarrow 
   \langle\langle \Box \phi_2 \rangle\rangle_k
   \end{array}
\right)
\right)\\

\end{array}
\]
\[\begin{array}{c|c}
    \phi & 1 \le i \le k  \\
   \hline

    \phi_1 \U \phi_2 &

    \begin{array}{c}
    \langle\langle \diamondsuit \phi_2 \rangle\rangle_i \iff
    \langle\langle \diamondsuit \phi_2 \rangle\rangle_{i-1} \lor
    ( \mathrm{InLoop}_i \land |[ \phi_2 ]|_i )
    \end{array} \\

   \phi_1 \R \phi_2 &

   \begin{array}{c}
   \langle\langle \Box \phi_2 \rangle\rangle_i \iff
   \langle\langle \Box \phi_2 \rangle\rangle_{i-1} \land
   ( \neg \mathrm{InLoop}_i \lor |[ \phi_2 ]|_i ) \\
   \end{array} \\
\end{array}
\]

\noindent {\em Last state constraints}:
\[
\begin{array}{c|c}
    \mathrm{Base} &
    \neg \mathrm{LoopExists} \Rightarrow \neg |[ \phi ]|_{k+1}
    \\
    \hline
    1 \le i \le k &
    l_i \Rightarrow ( |[ \phi ]|_{k+1} \iff |[ \phi ]|_{i})
\end{array}
\]

\noindent {\em First state constraints}:
\[\begin{array}{c|c}
    \phi &  \text{at 0}  \\ 
    \hline 
    \phi_1 \SNC \phi_2  &

    |[ \phi_1 \SNC \phi_2 ]|_0 \iff
    |[ \phi_2 ]|_0

    \\

   \phi_1 \TRG \phi_2 &

    |[ \phi_1 \TRG \phi_2 ]|_0 \iff
    |[ \phi_2 ]|_0

    \\

    \bullet \phi_1 &
    \neg |[ \bullet \phi_1 ]|_0 \\

    \bullet' \phi_1 &
    |[ \bullet' \phi_1 ]|_0 \\

\end{array}
\]

\medskip

\subsection*{Encoding of the metric operators}\label{mono-metric-stuff}

As before with the yesterday operator, for the mono-infinite encoding
of metric temporal operators we have to define, for all the metric
past operators, their duals w.r.t. negation.
Following the same notation used before, we will call
$\blacksquare'_{\sim t}$ the dual of $\blacksquare_{\sim t}$, and
$\blacklozenge'_{\sim t}$ the dual of $\blacklozenge_{\sim t}$.

By default, $\blacksquare_{\sim t}$, $\blacklozenge_{\sim t}$ are assumed to
be false when referring to time instants before 0, where  
$\blacksquare'_{\sim t}$, $\blacklozenge'_{\sim t}$ are assumed to
be true.

The semantics of the metric past dual operators is given by the
following formulae:
\[
\blacksquare_{= t} \phi \iff 
\blacklozenge_{= t} \phi \iff 
\neg \blacklozenge'_{= t}  \neg \phi \iff 
\neg \blacksquare'_{= t}  \neg \phi,
\]
\[ 
\blacksquare_{\le t} \phi \iff 
\neg \blacksquare'_{\le t} \neg \phi,
\qquad
\blacklozenge_{\le t} \phi \iff 
\neg \blacklozenge'_{\le t} \neg \phi.
\]








Next, we present the constraints of Sub-section \ref{metric-stuff}, modified
for a mono-infinite time structure.

\noindent {\em Metric constraints}:
\[
\begin{array}{c|c}
\phi & 0 \le j \le t-1 \\
	\hline 

    \diamondsuit_{=t} \phi, \  
    \Box_{\le t} \phi,  \ 
    \diamondsuit_{\le t} \phi &
	\MF{\phi}{j} \iff \bigvee_{i=1}^{k} l_i \land
	|[\phi]|_{i+\text{mod}(j,k-i+1)} \\
\end{array}
\]

\noindent {\em Temporal subformulae constraints}:
\[
\begin{array}{c|c}
    \phi  &  0 \le i \le k \\
    \hline

    \diamondsuit_{=t} \phi &

    |[ \diamondsuit_{=t} \phi ]|_i \iff |[ \phi ]|_{i+t},
    \text{ when } i+t \le k \\
&
    |[ \diamondsuit_{=t} \phi ]|_i \iff 
    \MF{\phi}{t+i-k-1},
    \text{ elsewhere } \\


    \Box_{\le t} \phi &
    \Box_{\le t} \phi \iff
    \bigwedge_{j=1}^{\text{min}(t,k-i)} |[\phi]|_{i+j} 
    \land
    \bigwedge_{j=k+1-i}^{t} \MF{\phi}{i+j-k-1}  \\

    \diamondsuit_{\le t} \phi &
    \diamondsuit_{\le t} \phi \iff
    \bigvee_{j=1}^{\text{min}(t,k-i)} |[\phi]|_{i+j} 
    \lor
    \bigvee_{j=k+1-i}^{t} \MF{\phi}{i+j-k-1}  \\

\end{array}
\]
\[
\begin{array}{ccc}

\begin{array}{c|c}
\phi &  0 \le i < t \\
	\hline 

  \blacklozenge_{= t} \phi & 
\neg |[ \blacklozenge_{= t} \phi ]|_i
\\

 \blacklozenge'_{= t} \phi & 
|[ \blacklozenge'_{= t} \phi ]|_i
\\

\blacksquare_{\le t} \phi & 
\neg |[ \blacksquare_{\le t} \phi ]|_i
\\

\blacksquare'_{\le t} \phi & 
|[ \blacksquare'_{\le t} \phi ]|_i
\\

  \blacklozenge_{\le t} \phi & 
\neg |[ \blacklozenge_{\le t} \phi ]|_0
\\
 
  \blacklozenge'_{\le t} \phi &
|[ \blacklozenge'_{\le t} \phi ]|_0 
\\
      \end{array}  

&
\qquad
&

\begin{array}{c|c}
 \phi    &  t \le i \le k+1 \\
    \hline

    \blacklozenge_{=t} \phi &
    |[ \blacklozenge_{=t} \phi ]|_i \iff |[ \phi ]|_{i-t}
 \\

    \blacklozenge'_{=t} \phi &
    |[ \blacklozenge'_{=t} \phi ]|_i \iff |[ \phi ]|_{i-t}
 \\

    \blacksquare_{\le t} \phi &
    \blacksquare_{\le t} \phi \iff
    \bigwedge_{j=1}^{t} |[\phi]|_{i-j}  \\

    \blacksquare'_{\le t} \phi &
    \blacksquare'_{\le t} \phi \iff
    \bigvee_{j=1}^{t} |[\phi]|_{i-j} 
     \\

    \blacklozenge_{\le t} \phi &
    \blacklozenge_{\le t} \phi \iff
    \bigvee_{j=1}^{t} |[\phi]|_{i-j} 
     \\

    \blacklozenge'_{\le t} \phi &
    \blacklozenge'_{\le t} \phi \iff
    \bigwedge_{j=1}^{t} |[\phi]|_{i-j}  \\

\end{array}

\end{array}
\]

\noindent {\em Past in Loop constraints}:
\[
\begin{array}{c|c}
   \phi  &  1 \le i \le k \\
    \hline

    \blacklozenge_{=t} \phi &

    l_i \Rightarrow
    \left(
    \begin{array}{c}
    \bigwedge_{j=1}^{\text{min}(t,i)} 
    (
        |[\phi]|_{i-j} 
	\iff
	|[\phi]|_{k-\text{mod}(j-1,k-i+1)}
    ) \land \\

    \bigwedge_{j=1+i}^{t}
    \left(
    \begin{array}{c}
    \neg |[\phi]|_{k-\text{mod}(j-1,k-i+1)}
    \end{array}
    \right)
  \end{array} 
  \right) \\

\\
    \blacklozenge'_{=t} \phi &

    l_i \Rightarrow
    \left(
    \begin{array}{c}
    \bigwedge_{j=1}^{\text{min}(t,i)} 
    (
        |[\phi]|_{i-j} 
	\iff
	|[\phi]|_{k-\text{mod}(j-1,k-i+1)}
    ) \land \\

    \bigwedge_{j=1+i}^{t}
    \left(
    \begin{array}{c}
    |[\phi]|_{k-\text{mod}(j-1,k-i+1)}
    \end{array}
    \right)
  \end{array} 
  \right) \\
\end{array}
\]
\[
\begin{array}{c|c}
    \phi &  1 \le i \le k \\
    \hline
   \blacksquare_{\le t} \phi &

   \mathrm{InLoop}_i \Rightarrow
    \left(
    
    |[\blacksquare_{\le t} \phi]|_i \iff
\left(
    \begin{array}{c}
	    \bigwedge_{j=0}^{\text{min}(k-i,t-1)}
	    (
	    \mathrm{InLoop}_{\text{max}(0,i-t+j)}
\lor 
	    |[\phi]|_{k-j}
	   ) \land \\
	    \bigwedge_{j=1}^{\text{min}(i,t)}
	    (
	    \neg \mathrm{InLoop}_{i-j}
\lor 
	    |[\phi]|_{i-j} 
	    ) 
    \end{array}

    \right)
    \right)

\\
  \blacksquare'_{\le t} \phi &

   \mathrm{InLoop}_i \Rightarrow
    \left(
    
    |[\blacksquare'_{\le t} \phi]|_i \iff
\left(
    \begin{array}{c}
	    \bigvee_{j=1}^{\text{min}(i,t)}
	    (
	    \mathrm{InLoop}_{i-j}
\land 
	    |[\phi]|_{i-j}
	   ) \lor \\
	    \bigvee_{j=0}^{\text{min}(k-i,t-1)}
	    (
	    \neg \mathrm{InLoop}_{\text{max}(0,i-t+j)}
\land 
	    |[\phi]|_{k-j} 
	    ) 
    \end{array}

    \right)
    \right)
\\

  \blacklozenge_{\le t} \phi &

   \mathrm{InLoop}_i \Rightarrow
    \left(
    
    |[\blacklozenge_{\le t} \phi]|_i \iff
\left(
    \begin{array}{c}
	    \bigvee_{j=1}^{\text{min}(i,t)}
	    (
	    \mathrm{InLoop}_{i-j}
\land 
	    |[\phi]|_{i-j}
	   ) \lor \\
	    \bigvee_{j=0}^{\text{min}(k-i,t-1)}
	    (
	    \neg \mathrm{InLoop}_{\text{max}(0,i-t+j)}
\land 
	    |[\phi]|_{k-j} 
	    ) 
    \end{array}

    \right)
    \right)
\\
   \blacklozenge'_{\le t} \phi &

   \mathrm{InLoop}_i \Rightarrow
    \left(
    
    |[\blacklozenge'_{\le t} \phi]|_i \iff
\left(
    \begin{array}{c}
	    \bigwedge_{j=0}^{\text{min}(k-i,t-1)}
	    (
	    \mathrm{InLoop}_{\text{max}(0,i-t+j)}
\lor 
	    |[\phi]|_{k-j}
	   ) \land \\
	    \bigwedge_{j=1}^{\text{min}(i,t)}
	    (
	    \neg \mathrm{InLoop}_{i-j}
\lor 
	    |[\phi]|_{i-j} 
	    ) 
    \end{array}

    \right)
    \right)
\\

\end{array}
\] 

\end{document}